\documentclass[journal]{IEEEtran}

\usepackage[dvips]{graphicx}
\usepackage[cmex10]{amsmath}
\usepackage{amsfonts}

\begin{document}

\title{Approximation to Distribution of Product of Random Variables Using Orthogonal Polynomials for Lognormal Density}

\author{Zhong~Zheng,~\IEEEmembership{Student~Member,~IEEE,}
        Lu~Wei,~\IEEEmembership{Student~Member,~IEEE,}
        Jyri~H\"{a}m\"{a}l\"{a}inen,~\IEEEmembership{Member,~IEEE,}
        and~Olav~Tirkkonen,~\IEEEmembership{Member,~IEEE}% <-this % stops a space
\thanks{The authors are with the Department of Communications and Networking, Aalto University, 00076 Aalto, Finland (e-mail: \{zhong.zheng, lu.wei, jyri.hamalainen, olav.tirkkonen\}@aalto.fi). J. H\"am\"al\"ainen is also with Ericsson Oulu R\&D Center.}}% <-this % stops a space

\maketitle

\begin{abstract}
%\boldmath
We derive a closed-form expression for the
orthogonal polynomials associated with the general lognormal
density. The result can be utilized to construct easily computable
approximations for probability density function of a product of random variables, when the considered variates are either independent or correlated. As an example, we have calculated the approximative distribution for the product of Nakagami-$m$ variables. Simulations indicate that accuracy of the proposed approximation is good with small cross-correlations under light fading condition.

\end{abstract}

\begin{IEEEkeywords}
Product of random variables, central limit theorem, lognormal
distribution, orthogonal polynomials, Nakagami-$m$.
\end{IEEEkeywords}

\section{Introduction}
\IEEEPARstart{S}{tatistical} properties of products of Random
Variables (RVs) are essential in performance analysis of contemporary wireless communication systems. For example, the fading
amplitude of multi-hop relaying systems follows the distribution of
the product of Nakagami-$m$ RVs~\cite{karagiannidis_2007}. In
addition, the cascaded-keyhole channel can be modeled using the product of individual keyhole channels~\cite{keyhole_2002}.

The exact Probability Density Function
(PDF) and Cumulative Distribution Function (CDF) of the product of
independent Beta, Gamma and Gaussian RVs can be represented in terms
of the Meijer-G function~\cite{springer_1970}. A more general
framework involving the Fox H-function was proposed
in~\cite{springer_1977} for the distribution of product of almost
any non-negative independent RVs. Although being theoretically interesting, these G- or H-function representations are difficult to evaluate on a low-complexity computing platform, where the calculation of the product statistics is required~\cite{simon_2004}. %For example, when a transmitter (base station or mobile station) only knows the distribution of the fading channel, the rate adaptation is based on the channel statistics~\cite{karagiannidis_2007,chen_2012}.
The numerical calculations of G- and H-functions require a numerical solution of contour integrals or a construction of look-up tables that cover all parameter combinations. Both approaches set stringent requirements for the considered platform either in the computing capability or the memory capacity. Recently, Ahmed {\it et al.}~\cite{ahmed_2011} employed the Mellin transform and the residue theorem to derive the PDF of a product of {\it i.i.d.} Nakagami-$m$ RVs as an infinite series. By exploiting the structure of the Mellin transforms, the authors in~\cite{lu_2011,chen_2012} proposed approximations to the PDFs and CDFs of products of independent Rayleigh, Gamma, Nakagami-$m$ and Gaussian RVs. However, when the RVs in product are correlated, exact distributions are known only in some special cases (e.g. bivariate Nakagami-$m$ in~\cite{nakagami_1957}). For arbitrary number of RVs with generic correlations, none of the aforementioned methodologies are suitable to give a tractable solution. This is the main issue to be addressed in this paper.

Motivated by the Central Limit Theorem (CLT), the approximation for the product of RVs can be constructed by using a lognormal density and its associated orthogonal polynomials. We apply the approach of~\cite{provost_2009} where the first few moments of the product and the approximated density function are set equal. The resulting approximative distribution involves only a finite sum of polynomials and the elementary lognormal density function. Therefore, implementing the proposed expressions on practical computing platforms is straightforward compared with the G- or H-function representations. We note that the proposed framework is suitable for cases of both independent and correlated variables provided that moments of a product can be computed in closed-form.

In this paper we deduce a closed-form expression for the orthogonal polynomials associated with a general lognormal density. The result is obtained by taking advantage of the determinant representation of orthogonal polynomials. To give a specific example, we derive an approximative PDF and CDF for the product of both independent and correlated \mbox{Nakagami-$m$} RVs. Numerical results are compared with simulations in terms of the Complementary CDF (CCDF). The overall approximation accuracy is measured by the Mean Square Error (MSE) between the approximative CDF and the empirical CDF. The orthogonal polynomials associated with a standard lognormal density\footnote{When the corresponding Gaussian distribution is zero mean and unit variance.} were first developed in~\cite{ernst_2011}, which is a special case of our result. It is worth of noticing that our results are not straightforward extensions of the results presented in~\cite{ernst_2011} but require new methodological elements.

\section{Orthogonal Polynomials Associated with General Lognormal Density}\label{sec_op}

Assume that $Y$ is a Gaussian RV with mean $\mu$ and variance $\sigma^2$. Then $X=e^Y$ follows the lognormal distribution
\begin{equation}
f_{\textrm{LN}}(x)=\frac{1}{x\sqrt{2\pi\sigma^2}}e^{-\frac{(\log x-\mu)^2}{2\sigma^2}},\quad x\in(0,\infty)
\end{equation}
and the $i$-th moment, $\nu_i$ ($i\in\mathbb{N}$), of $X$ is given by
\begin{equation}
\nu_i=\int_0^{\infty}x^i f_{\textrm{LN}}(x)\,\mathrm{d}x=e^{i\mu+\frac{1}{2}i^2\sigma^2}\label{eq_moment}.
\end{equation}

Let $\pi_n(x)$ be the $n$-th degree polynomial
\begin{equation}
\pi_n(x)=\sum_{k=0}^n c_{n,k}x^k,\label{eq_p}
\end{equation}
where $c_{n,n}\neq 0$. The polynomials
$\{\pi_n(x)\}$ are said to be orthogonal with respect to the general lognormal
density if
\begin{equation}\label{eq:OrthFac}
\int_0^\infty
\pi_j(x)\pi_k(x)f_{\textrm{LN}}(x)\mathrm{d}x=h_j\delta_{jk},\quad
j,k\in\mathbb{N},
\end{equation}
where $h_j=\sum_{i=0}^j\sum_{k=0}^j c_{j,i} c_{j,k} \nu_{i+k}$ is a normalization factor. The symbol $\delta_{jk}$ is the Kronecker delta symbol, which is defined as $\delta_{jk}=1$ if $j=k$ and zero otherwise.

Due to~(\ref{eq_moment}) any arbitrary moment of $f_{\textrm{LN}}(x)$
exists. Thus, functions $x^i$ ($i\in\mathbb{N}$) belong to the space of square integrable functions with respect to the weight function $f_{\textrm{LN}}(x)$. There exists a unique set of orthogonal polynomials $\{\pi_n(x)\}$ which
admits the explicit determinant representation $\pi_n(x)=(\Delta_n)^{-1}\Delta_n(x)$~\cite[Th.
2.1.1]{szego_1939},
%\begin{equation}
%\pi_n(x)=(\Delta_n)^{-1}\Delta_n(x),\label{eq_op}
%\end{equation}
where $\Delta_n(x)$ denotes the $n$-th degree polynomial
\begin{equation}
\Delta_n(x)=\left |
\begin{array}{ccccc}
\nu_0 & \cdots & \nu_{n-1} & \nu_n\\
\vdots & \ddots & \vdots & \vdots\\
\nu_{n-1}& \cdots & \nu_{2n-2} & \nu_{2n-1}\\
1 & \cdots & x^{n-1} & x^n
\end{array}
\right |,\label{eq_op_p1}
\end{equation}
and the constant $\Delta_n$ is formed by deleting the last row and column from
$\Delta_n(x)$, i.e. $\Delta_n=\left|\nu_{i+j}\right|_{i,j=0,\cdots,n-1}$
with $\Delta_0=1$. The determinant (\ref{eq_op_p1}) can be expanded using the cofactors with respect to the last row. Then $\pi_n(x)$ becomes
\begin{equation}
\pi_n(x)=\sum_{k=0}^n (-1)^{n+k}\frac{\Delta_{n,k}}{\Delta_n}x^k\label{eq_op_cofactor},
\end{equation}
where the cofactor $\Delta_{n,k}$ with respect to $x^k$ is obtained
by removing the $(k+1)$-th column and the last row from
(\ref{eq_op_p1}), $\Delta_{0,0}=0$ and $\Delta_{1,0}=1$. After comparing (\ref{eq_op_cofactor}) with (\ref{eq_p}), we obtain
\begin{equation}
c_{n,k}=(-1)^{n+k}\frac{\Delta_{n,k}}{\Delta_n}, \quad k=0,\cdots,n.\label{eq_cnk}
\end{equation}
Let us calculate an explicit expression for $c_{n,k}$. First, by the definition of $\Delta_{n,k}$ and $\Delta_n$ it is observed that $c_{n,n}=1$, i.e. the orthogonal polynomials ($\ref{eq_op_cofactor}$) are monic\footnote{Monic polynomial is defined as the polynomial where the coefficient of highest degree term is unity.}. Furthermore, after inserting (\ref{eq_moment}) into $\Delta_{n,k}$ and denoting $q=e^{\sigma^2}$, we obtain (see Appendix for details)
\begin{equation}
\Delta_{n,k}=E(n)\frac{\nu_n\prod_{i=0}^{n-1}\prod_{j=i+1}^n(q^j-q^i)}{\nu_k\prod_{j=k+1}^n(q^j-q^k)\prod_{i=0}^{k-1}(q^k-q^i)},\label{eq_cofactor_expand}
\end{equation}
where $E(n)=e^{n(n-1)\mu+\frac{\sigma^2}{6}n(2n^2-3n+1)}$. For
$k=n$, $\Delta_n=\Delta_{n,n}$ and (\ref{eq_cofactor_expand})
becomes
\begin{equation}
\Delta_n=E(n) \prod_{i=0}^{n-2}\prod_{j=i+1}^{n-1}\left(q^j-q^i\right).\label{eq_op_p2_expand}
\end{equation}
After substituting (\ref{eq_cofactor_expand}) and (\ref{eq_op_p2_expand})
into (\ref{eq_cnk}) we find that %$c_{n,k}=(-1)^{n+k}e^{(n-k)\mu}q^{(n-1/2)(n-k)} {n\brack k}_q$
\begin{equation}
c_{n,k}=(-1)^{n+k}e^{(n-k)\mu}q^{(n-1/2)(n-k)} {n\brack k}_q\label{eq_cnk2},
\end{equation}
where ${n\brack k}_q=\frac{(1-q^n)(1-q^{n-1})\cdots(1-q^{n-k+1})}{(1-q^k)(1-q^{k-1})\cdots(1-q)}$
%\begin{equation}
%{n\brack
%k}_q=\frac{(1-q^n)(1-q^{n-1})\cdots(1-q^{n-k+1})}{(1-q^k)(1-q^{k-1})\cdots(1-q)}
%\end{equation}
is the generalized binomial coefficient.

\section{Approximation to the Density of Product of Nakagami-$m$ Random Variables}\label{sec_approx}

\subsection{General Framework}

According to the CLT, the distribution of a product of RVs can be
approximated by a lognormal distribution when the number of RVs is
large. Motivated by this, we choose $f_{\mathrm{LN}}(x)$ as an
initial approximation for the distribution of a product of RVs in the
context of the moment based density approximation, which is derived in~\cite{provost_2009}. Let $M(k)$ denote the $k$-th moment of a density function $f(x)$, the approximation to $f(x)$ reads
\begin{equation}
f(x)\simeq f_{\mathrm{LN}}(x)\sum_{i=0}^N\eta_i\pi_i(x)\label{eq_approximant},
\end{equation}
where the constant $\eta_{i}=\frac{1}{h_{i}}\sum_{k=0}^i c_{i,k}M(k)$. Note that the first $N$ moments of the approximation (\ref{eq_approximant}) are matched with the corresponding moments of $f(x)$~\cite{provost_2009}.

Alternatively, (\ref{eq_approximant}) can be rearranged by combining the coefficients of $\pi_i(x)$ with the same power of $x$, resulting in $f(x)\simeq f_{\mathrm{LN}}(x)\sum_{i=0}^N\xi_i x^{i}$
%\begin{equation}\label{eq:PowForm}
%f(x)\simeq f_{\mathrm{LN}}(x)\sum_{i=0}^N\xi_i x^{i},
%\end{equation}
where $\xi_j=\sum_{k=j}^N c_{k,j}\eta_k$. After direct integration
%of~(\ref{eq:PowForm}) 
the approximated CDF attains the form
\begin{equation}\label{eq:cdf}
F(x)\simeq \sum_{i=0}^N\xi_i \nu_i\Phi\left(\frac{\log(x)-\mu}{\sigma}-i\sigma\right),
\end{equation}
where $\Phi(\cdot)$ is the CDF of a standard normal RV. The CCDF $\bar{F}(x)=1-F(x)$ is approximated by replacing $F(x)$ with the expression~(\ref{eq:cdf}).

It is a typical situation that the moments of certain RV are
relatively easy to obtain whereas its exact distribution is
difficult to calculate or unavailable~\cite{2006Ha}. The proposed
approximations~(\ref{eq_approximant}) and~(\ref{eq:cdf}) are particularly useful in this situation, as will be shown in the next subsection.

\subsection{Moment Based Approximation to Distribution of Product of Nakagami-$m$ RVs}

Let RV $P=\prod_{i=1}^K R_i$ be a product of $K$
Nakagami-$m$ RVs each with the PDF
\begin{equation}
f_{R_i}(x)\!=\!\frac{2m_i^{m_i}x^{2m_i-1}}{\Omega_i^{m_i}\Gamma(m_i)}\!\exp\left(-\frac{m_i}{\Omega_i}x^2\right),~~x\in(0,\infty).
\end{equation}
We first consider the case when the variables $R_i$ are correlated with each other. In the literature, there exists different representations for the joint PDF of multivariate Nakagami-$m$ RVs, which are either limited
in parameter values or cross-correlation structures. In this paper,
we adopt a recent result derived in~\cite{beaulieu_2011}, which gives the
joint PDF as a single integral. In the proposed approach, the RVs $R_i$ are assumed to have the same fading parameter $m$ and the joint PDF is valid for
integer and half-integer values of $m$. The power cross-correlation
coefficient between $R_i^2$ and $R_j^2$ is of the form
\begin{equation}
\rho_{R_i^2,R_j^2}=\frac{\mathbb{E}[R_i^2
R_j^2]-\mathbb{E}[R_i^2]\mathbb{E}[R_j^2]}{\sqrt{\mbox{Var}[R_i^2]\mbox{Var}[R_j^2]}}=\lambda_i^2\lambda_j^2.
\end{equation}
Note that the derived approximation framework is not limited by the
above multivariate model.

Based on the joint PDF~\cite[eq. (20)]{beaulieu_2011}, the $k$-th moment of
the RV $P$ is calculated by using~\cite[eq.
(6.643/2)]{gradshteyn_2007} as
\begin{align}\label{eq:ProMoCor}
\hspace{-1ex}M(k)&=\frac{\Gamma(m+k/2)^K}{m^{Kk/2}\Gamma(m)^{K+1}}\prod_{i=1}^K\left[\Omega_i(1-\lambda_i^2)\right]^{\frac{k}{2}}\nonumber\\
&\times\int_0^{\infty}t^{m-1}e^{-t}\prod_{i=1}^K{\mbox{$_1$F$_1\!$}}\left(-\frac{k}{2},m,\frac{\lambda_i^2 t}{\lambda_i^2-1}\right)\,dt,
\end{align}
where ${\mbox{$_1$F$_1\!$}}\left(a,b,c\right)$ denotes the Kummer
confluent hypergeometric function. The parameters $\mu$ and
$\sigma^2$ of $f_{\mathrm{LN}}(x)$ can be obtained by equating them
with the mean and variance of $\log(P)$ respectively as
\begin{align}
\mu
&=\sum_{i=1}^K\mathbb{E}[\log(R_i)]=\frac{1}{2}\sum_{i=1}^K\left[\Psi_0(m)-\log\left(\frac{m}{\Omega_i}\right)\right],\label{eq:meanCor}\\
\sigma^2\! &=\! \sum_{i=1}^K\mbox{Var}[\log(R_i)]+2\sum_{i<j} \mbox{Cov}[\log(R_i),\log(R_j)]\nonumber\\
&\hspace{-2ex}=\!\frac{1}{4}\!\sum_{i=1}^K\! \Psi_1(m)\!+\!2\!\sum_{i<j}\!\left(\!\int_0^{\infty}\!
\frac{I_i(t)I_j(t)t^{m-1}}{4\Gamma(m)e^{t}}\,dt\!-\!\zeta_i\zeta_j\!\right),\label{eq:varCor}
\end{align}
where $\Psi_n(x)$ is the polygamma function that is defined as the $(n+1)$-th derivative of the logarithm of $\Gamma(x)$, and $\zeta_i=\frac{1}{2}\left[\Psi_0(m)-\log\left(\frac{m}{\Omega_i}\right)\right]$. Here the function $I_i(t)$ is
\begin{equation*}
I_i(t)\!=\!\Psi_0(m)\!+\!\log\!\left(\!\frac{\Omega_i(1-\lambda_i^2)}{m}\!\right)\!-\!{\mbox{$_1$F$_1\!$}}^{'}\left.\!\left(\!a,m,\frac{\lambda_i^2
t}{\lambda_i^2-1}\!\right)\right|_{a=0},
\end{equation*}
where ${\mbox{$_1$F$_1\!$}}^{'}\left(a,b,c\right)$ refers to the
derivative of ${\mbox{$_1$F$_1\!$}}\left(a,b,c\right)$ with respect
to the parameter $a$.

Next consider the case where the RVs $R_{i}$ are independent. We notice that with $\lambda_i=0$, the joint PDF~\cite[eq. (20)]{beaulieu_2011} is in the form of a product of individual PDFs of $R_i$ ($i=1,2,\cdots,K$), which implies statistical independence. Thus, we let $\lambda_i=0$ and calculate the moments of $P$ from~(\ref{eq:ProMoCor})
\begin{equation}\label{eq_mo_indi}
M(k)=\prod_{i=1}^K\frac{\Gamma(m+k/2)}{\Gamma(m)}\left(\frac{\Omega_i}{m}\right)^{k/2}.
\end{equation}
While the calculation of $\mu$ is not affected by the independency and it is given by expression~(\ref{eq:meanCor}), the variance $\sigma^2$ is reduced to the form
\begin{equation}\label{eq_var_indi}
\sigma^2=\frac{1}{4}\sum_{i=1}^K \Psi_1(m)
\end{equation}
with all the covariances $\mathrm{Cov}[\log(R_i),\log(R_j)]$ in~(\ref{eq:varCor}) equal to zero. It is noted that (\ref{eq_mo_indi}) and (\ref{eq_var_indi}) agree with the results given by~\cite{karagiannidis_2007}.

\begin{figure}[!t]
\centering
\includegraphics[width=3.2in]{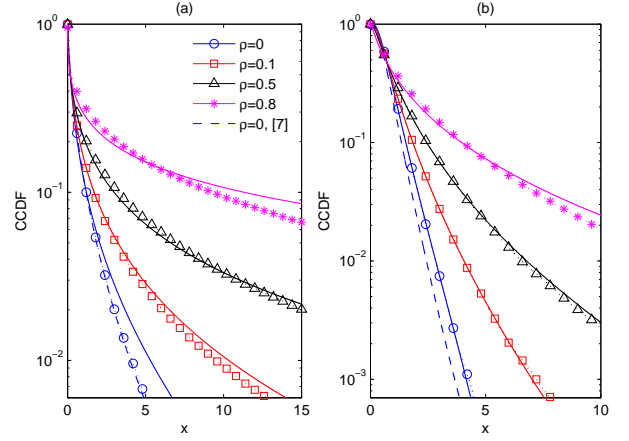}
\caption{CCDF of products of six Nakagami-$m$ RVs when $\Omega=1$ and $N=16$. (a) $m=1$; (b) $m=4$. Solid lines: approximative PDFs; markers: simulated PDFs.} \label{fig_ccdf}
\end{figure}

\begin{table*}[!t]
% increase table row spacing, adjust to taste
%\renewcommand{\arraystretch}{1.3}
\caption{Mean square error $\epsilon^2$ of approximative distribution}
\label{tab_mse}
\centering
% Some packages, such as MDW tools, offer better commands for making tables
% than the plain LaTeX2e tabular which is used here.
\begin{tabular}{|l|c|c|c|c|c|c|c|c|c|c|c|c|c|c|c|c|c|c|c|c|}
\hline
 \hspace{7ex}{$K$} 						& 2 			& 4 			& 6 			& 8 			& 10 			& 12 			& 14 			& 16 			& 18 			& 20\\\hline
$m=1$, $\rho=0$ 	& 1.14e-3 & 1.09e-3 & 6.28e-4 & 3.78e-4 & 2.72e-4 & 2.02e-4	& 1.67e-4	& 1.45e-4 & 1.28e-4 & 1.14e-4\\\hline
$m=1$, $\rho=0.1$& 1.06e-3 & 5.00e-4 & 1.05e-4 & 1.66e-5 & 1.73e-5 & 5.35e-5 & 1.03e-4 & 1.72e-4 & 2.40e-4 & 3.21e-4\\\hline
$m=1$, $\rho=0.5$& 9.99e-4 & 1.68e-4 & 2.10e-5 & 9.54e-5 & 2.07e-4 & 3.18e-4 & 4.16e-4 & 5.36e-4 & 6.24e-4 & 6.98e-4\\\hline
$m=1$, $\rho=0.8$& 2.17e-3 & 9.16e-4 & 7.72e-4 & 7.41e-4 & 7.28e-4 & 7.21e-4 & 7.09e-4 & 7.15e-4 & 7.13e-4 & 7.16e-4\\\hline
$m=4$, $\rho=0$ 	& 8.13e-6 & 2.29e-5 & 2.43e-5 & 1.64e-5 & 2.24e-5 & 3.15e-5	& 4.34e-5	& 4.78e-5 & 5.14e-5 & 5.02e-5\\\hline
$m=4$, $\rho=0.1$& 8.31e-6 & 1.16e-5 & 2.65e-6 & 2.33e-6 & 1.77e-5 & 4.95e-5 & 8.93e-5 & 1.15e-4 & 1.31e-4 & 1.47e-4\\\hline
$m=4$, $\rho=0.5$& 7.03e-6 & 1.34e-5 & 7.44e-6 & 9.73e-6 & 1.74e-5 & 2.48e-5 & 3.36e-5 & 4.21e-5 & 4.99e-5 & 5.49e-5\\\hline
$m=4$, $\rho=0.8$& 3.58e-5 & 3.20e-4 & 4.63e-4 & 3.03e-4 & 2.27e-4 & 2.12e-4 & 2.09e-4 & 2.11e-4 & 2.09e-4 & 2.06e-4\\\hline
\end{tabular}
\end{table*}

\section{Numerical Results}
In Fig.~\ref{fig_ccdf}, we compare simulations with the approximative CCDFs for the product of six Nakagami-$m$ RVs with parameters $m=1,\ 4$ and $\Omega=1$, using orthogonal polynomials of up to 16-th degree. Here we consider equal cross-correlations, i.e. $\rho_{R_i^2,R_j^2}=\rho$~($i\neq j$), and calculate the parameters of the approximation using~(\ref{eq:ProMoCor}),~(\ref{eq:meanCor}) and~(\ref{eq:varCor})  with $\rho=0.1,\ 0.5\mathrm{\ and\ }0.8$ for correlated Nakagami-$m$ RVs. In case of independent RVs (with $\rho=0$), equations~(\ref{eq_mo_indi}), (\ref{eq:meanCor}) and (\ref{eq_var_indi}) are applied. When $m=1$ the RVs $R_i$ are Rayleigh distributed, while with $m=4$ the $R_i$ are approximately Rician distributed with the Rician $\kappa$-factor given by $\kappa=6.46$~\cite{simon_2004}. For each simulated CCDF curve, we generate $10^6$ realizations of the \mbox{Nakagami-$m$} RVs $R_i$ ($i=1,\cdots,6$) using Sim's method \cite{sim_1993}. As a comparison, we also plot the approximative CCDFs calculated from~\cite[eq. (36)]{chen_2012} for the cases with independent Nakagami-$m$ RVs. Fig.~\ref{fig_ccdf}(b) shows that differences between the approximative CCDFs and simulations are less than $10^{-3}$ when $\rho=0\ \mathrm{and}\ 0.1$, and less than $10^{-2}$ when $\rho=0.5$. In addition, when $m=4$ and $\rho=0$, the proposed approximation yields improved accuracy compared with the results given in~\cite{chen_2012}. However, as $m=1$, the approximations with the same correlation coefficients are noticeably inaccurate due to slow convergence of the series in~(\ref{eq:cdf}). Note that in both cases the approximative CCDFs with $\rho=0.8$ deviate from the simulations since the CLT fails under high correlation condition.

Table~\ref{tab_mse} summarizes the MSE $\epsilon^2=\int_0^{\infty}|F^*(x)-F(x)|^2\,dF(x)$ between the approximation $F(x)$ and the empirical distribution function $F^*(x)$ of the product $P$ as a function of the number of RVs $K$. When $K$ is small~($K\le 10$), Table~\ref{tab_mse} shows that the approximated CDF achieves considerably smaller MSE in low-correlation cases ($\rho=0.1\ \mathrm{and}\ 0.5$) compared with the independent case ($\rho=0$). As $K$ increases, the accuracy for the independent case becomes superior to those of correlated ones. When $m=4$, all approximations are much more accurate compared with the corresponding Rayleigh cases ($m=1$), and the MSEs $\epsilon^2$, except for those where $K>16$ and $\rho=0.1$, is less than $10^{-4}$ in case of independent RVs or when correlation is low.

Although the accuracies of the approximations~(\ref{eq_approximant}) and~(\ref{eq:cdf}) should be improved as $N$ increases, analysis of the exact improvement is difficult due to the multiple nested finite summations involved. However, numerical results indicate that the $i$-th term in the summation~(\ref{eq:cdf}) converges quickly to zero since the coefficient $\xi_i\nu_i$ converges to zero and the CDF $\Phi\left(\frac{\log(x)-\mu}{\sigma}-i\sigma\right)$ is bounded. The value $N=16$ is shown to be sufficient to yield a stable CDF approximation.

\section{Conclusion}

Knowledge of the distributions of product of random variables is
important for understanding the performance of various communication
systems. In this work, we first derived a closed-form expression for
the orthogonal polynomials associated with general lognormal
density. The derived result was subsequently applied in
approximating the distribution of the product of random variables.
%in a moment based framework
As an example, we calculated closed-form approximations for the
distributions of product of Nakagami-$m$ variates. Under light fading conditions, the resulting expressions achieve a good trade-off between computation complexity and approximation accuracy for small cross-correlations. It is still an open problem to find a closed-form (approximated) distribution for the product of correlated Nakagami-$m$ RVs under deep fading condition.

\appendix
We note that the determinant
\begin{equation}
\Delta_{n,k} = \left | e^{(i+j)\mu+\frac{(i+j)^2}{2}\sigma^2}
\right|_{i=0,\cdots,n-1;\; j=0,\cdots,n;\; j\neq k}
\end{equation}
remains unchanged by factoring out the term $e^{i\mu+i^2\sigma^2/2}$
from the $i$-th row  and the term $e^{j\mu+j^2\sigma^2/2}$ from the $j$-th
column. Namely, we have
\begin{align}
\Delta_{n,k} & = E(n)\frac{\nu_n}{\nu_k}\left|e^{ij\sigma^2}\right|_{i=0,\cdots,n-1;\;j=0,\cdots,n;\;j\neq k}\\
&=E(n)\frac{\nu_n\prod_{i=0}^{n-1}\prod_{j=i+1}^n(q^j-q^i)}{\nu_k\prod_{j=k+1}^n(q^j-q^k)\prod_{i=0}^{k-1}(q^k-q^i)},
\end{align}
\noindent where $E(n)=e^{n(n-1)\mu+\frac{\sigma^2}{6}n(2n^2-3n+1)}$ and the
second equality follows from the fact that
$\left|e^{ij\sigma^2}\right|$ is a Vandermonde determinant.

\section*{Acknowledgment}
This work was supported in part by the Academy of Finland (grant 133562) and in part by the Finnish Funding Agency for Technology and Innovation (TEKES).

% Can use something like this to put references on a page
% by themselves when using endfloat and the captionsoff option.
\ifCLASSOPTIONcaptionsoff
  \newpage
\fi

\end{document}